\documentclass[10pt,letterpaper]{article}
\usepackage{opex3}
\usepackage{times}
\usepackage[]{cite}
\usepackage{SIunits}
\begin{document}

\title{Depolarized guided acoustic wave Brillouin scattering in hollow-core photonic crystal fibers}

\author{Wenjia Elser n\'{e}e Zhong$^{1,2,*}$, Birgit Stiller$^{1,2}$, Dominique Elser$^{1,2}$, Bettina Heim$^{1,2}$, Christoph Marquardt$^{1,2}$, and Gerd Leuchs$^{1,2}$}

\address{
$^{1}$ Max Planck Institute for the Science of Light, G\"unther-Scharowsky-Str.\ 1, bld. 24\\ D-91058 Erlangen, Germany \\
$^{2}$ Institute of Optics, Information and Photonics, University of Erlangen-Nuremberg,\\ Staudtstr. 7/B2, D-91058 Erlangen, Germany \\
$^{*}$ wenjia.zhong@mpl.mpg.de
}

\begin{abstract}
By performing quantum-noise-limited optical heterodyne detection, we observe polarization noise in light after propagation through a hollow-core photonic crystal fiber (PCF). We compare the noise spectrum to the one of a standard fiber and find an increase of noise even though the light is mainly transmitted in air in a hollow-core PCF. Combined with our simulation of the acoustic vibrational modes in the hollow-core PCF, we are offering an explanation for the polarization noise with a variation of guided acoustic wave Brillouin scattering (GAWBS). Here, instead of modulating the strain in the fiber core as in a solid core fiber, the acoustic vibrations in hollow-core PCF influence the effective refractive index by modulating the geometry of the photonic crystal structure. This induces polarization noise in the light guided by the photonic crystal structure.
\end{abstract}
\ocis{
(060.4005)   Microstructured fibers;
(060.5295)   Photonic crystal fibers;
(060.2270)   Fiber characterization;
(110.5125)   Photoacoustics;
(290.5830)   Scattering, Brillouin;
(290.2558)   Forward scattering;
} 

\bibliographystyle{osajnl}
\bibliography{OE2011}

\begin{thebibliography}{10}
\newcommand{\enquote}[1]{``#1''}

\bibitem{Shelby1985}
R.~M. Shelby, M.~D. Levenson, and P.~W. Bayer, \enquote{{Guided acoustic-wave
  Brillouin scattering},} Physical Review B \textbf{31}, 5244--5252 (1985).

\bibitem{Shelby1985a}
R.~Shelby, M.~Levenson, and P.~Bayer, \enquote{{Resolved forward Brillouin
  scattering in optical fibers},} Physical Review Letters \textbf{54}, 939--942
  (1985).

\bibitem{Nishizawa1995}
N.~Nishizawa, S.~Kume, M.~Mori, T.~Goto, and A.~Miyauchi,
  \enquote{{Experimental analysis of guided acoustic wave Brillouin scattering
  in PANDA fibers},} Journal of the Optical Society of America B \textbf{12},
  1651 (1995).

\bibitem{Nishizawa1996}
N.~Nishizawa, S.~Kume, M.~Mori, T.~Goto, and A.~Miyauchi,
  \enquote{{Characteristics of guided acoustic wave Brillouin scattering in
  polarization maintaining fibers},} Optical Review \textbf{3}, 29--33 (1996).

\bibitem{Laude2005}
V.~Laude, A.~Khelif, S.~Benchabane, M.~Wilm, T.~Sylvestre, B.~Kibler,
  A.~Mussot, J.~Dudley, and H.~Maillotte, \enquote{{Phononic band-gap guidance
  of acoustic modes in photonic crystal fibers},} Physical Review B
  \textbf{71}, 1--6 (2005).

\bibitem{Shibata2006}
N.~Shibata, A.~Nakazono, N.~Taguchi, and S.~Tanaka, \enquote{{Forward Brillouin
  scattering in holey fibers},} IEEE Photonics Technology Letters \textbf{18},
  412--414 (2006).

\bibitem{Elser2006}
D.~Elser, U.~L. Andersen, A.~Korn, O.~Gl\"{o}ckl, S.~Lorenz, C.~Marquardt, and
  G.~Leuchs, \enquote{{Reduction of guided acoustic wave Brillouin scattering
  in photonic crystal fibers},} Physical Review Letters \textbf{97}, 133901
  (2006).

\bibitem{Beugnot2007}
J.-C. Beugnot, T.~Sylvestre, H.~Maillotte, G.~M\'{e}lin, and V.~Laude,
  \enquote{{Guided acoustic wave Brillouin scattering in photonic crystal
  fibers.}} Optics Letters \textbf{32}, 9--17 (2007).

\bibitem{McElhenny2011}
J.~E. McElhenny, R.~K. Pattnaik, and J.~Toulouse, \enquote{{Dependence of
  frequency shift of depolarized guided acoustic wave Brillouin scattering in
  photonic crystal fibers},} Journal of Lightwave Technology \textbf{29},
  200--208 (2011).

\bibitem{Stiller_2011}
B.~Stiller, M.~Delqu\'{e}, J.-C. Beugnot, M.~W. Lee, G.~M\'{e}lin,
  H.~Maillotte, V.~Laude, and T.~Sylvestre, \enquote{{Frequency-selective
  excitation of guided acoustic modes in a photonic crystal fiber},} Optics
  Express \textbf{19}, 7689--7694 (2011).

\bibitem{Shelby1986}
R.~M. Shelby, M.~D. Levenson, S.~H. Perlmutter, R.~G. DeVoe, and D.~F. Walls,
  \enquote{{Broad-band parametric deamplification of quantum noise in an
  optical fiber},} Physical Review Letters \textbf{57}, 691--694 (1986).

\bibitem{Dong2008}
R.~Dong, J.~Heersink, J.~F. Corney, P.~D. Drummond, U.~L. Andersen, and
  G.~Leuchs, \enquote{{Experimental evidence for Raman-induced limits to
  efficient squeezing in optical fibers},} Optics Letters \textbf{33}, 116
  (2008).

\bibitem{Lorenz2001}
S.~Lorenz, C.~Silberhorn, N.~Korolkova, R.~S. Windeler, and G.~Leuchs,
  \enquote{Squeezed light from microstructured fibres: towards free-space
  quantum cryptography,} Applied Physics B: Lasers and Optics \textbf{73}, 855
  -- 859 (2001).

\bibitem{Josip2007}
J.~Milanovic, J.~Heersink, C.~Marquardt, A.~Huck, U.~L. Andersen, and
  G.~Leuchs, \enquote{Polarization squeezing with photonic crystal fibers,}
  Laser Physics \textbf{17}, 559--566 (2007).

\bibitem{Lodewyck2007}
J.~Lodewyck, M.~Bloch, R.~Garc\'{\i}a-Patr\'{o}n, S.~Fossier, E.~Karpov,
  E.~Diamanti, T.~Debuisschert, N.~Cerf, R.~Tualle-Brouri, S.~McLaughlin, and
  P.~Grangier, \enquote{{Quantum key distribution over 25km with an all-fiber
  continuous-variable system},} Physical Review A \textbf{76} (2007).

\bibitem{Wittmann2010}
C.~Wittmann, J.~F\"{u}rst, C.~Wiechers, D.~Elser, H.~H\"{a}seler,
  N.~L\"{u}tkenhaus, and G.~Leuchs, \enquote{{Witnessing effective entanglement
  over a 2km fiber channel},} Optics Express \textbf{18}, 4499 (2010).

\bibitem{Elser2007}
D.~Elser, C.~Wittmann, U.~L. Andersen, O.~Gl\"ockl, S.~Lorenz, C.~Marquardt,
  and G.~Leuchs, \enquote{Guided acoustic wave {B}rillouin scattering in
  photonic crystal fiber,} Journal of Physics: Conference Series \textbf{92},
  012108 (2007).

\bibitem{Knight1998}
J.~C. Knight, \enquote{{Photonic band gap guidance in optical fibers},} Science
  \textbf{282}, 1476--1478 (1998).

\bibitem{Cregan1999}
R.~F. Cregan, \enquote{{Single-mode photonic band gap guidance of light in
  air},} Science \textbf{285}, 1537--1539 (1999).

\bibitem{RussellPCF}
P.~S.~J. Russell, \enquote{Photonic crystal fibers,} Science \textbf{299},
  358--362 (2003).

\bibitem{Ouzounov2003}
D.~G. Ouzounov, F.~R. Ahmad, D.~M\"uller, N.~Venkataraman, M.~T. Gallagher,
  M.~G. Thomas, J.~Silcox, K.~W. Koch, and A.~L. Gaeta, \enquote{Generation of
  megawatt optical solitons in hollow-core photonic band-gap fibers,} Science
  \textbf{301}, 1702--1704 (2003).

\bibitem{Couny2007b}
F.~Couny, F.~Benabid, and P.~S. Light, \enquote{Subwatt threshold cw {R}aman
  fiber-gas laser based on {H}$_{2}$-filled hollow-core photonic crystal
  fiber,} Physical Review Letters \textbf{99}, 143903 (2007).

\bibitem{Bykov2015}
D.~S. Bykov, O.~A. Schmidt, T.~G. Euser, and P.~S.~J. Russell, \enquote{Flying
  particle sensors in hollow-core photonic crystal fibre,} Nature Photonics
  \textbf{9}, 461–465 (2015).

\bibitem{Benabid2002}
F.~Benabid, J.~Knight, and P.~S.~J. Russell, \enquote{Particle levitation and
  guidance in hollow-core photonic crystal fiber,} Optics Express \textbf{10},
  1195--1203 (2002).

\bibitem{Ana2013}
A.~M. Cubillas, S.~Unterkofler, T.~G. Euser, B.~J.~M. Etzold, A.~C. Jones,
  P.~J. Sadler, P.~Wasserscheid, and P.~S.~J. Russell, \enquote{Photonic
  crystal fibres for chemical sensing and photochemistry,} Chemical Society
  Reviews \textbf{42}, 8629--8648 (2013).

\bibitem{KaFai2013}
K.~F. Mak, J.~C. Travers, P.~H\"olzer, N.~Y. Joly, and P.~S.~J. Russell,
  \enquote{Tunable vacuum-{UV} to visible ultrafast pulse source based on
  gas-filled {K}agome-{PCF},} Optics Express \textbf{21}, 10942--10953 (2013).

\bibitem{Zhong2007}
{W. Zhong}, {Ch. Marquardt}, {G. Leuchs}, {U. L. Andersen}, {P. Light}, {F.
  Couny}, and {F. Benabid}, \enquote{{Squeezing by self induced transparency in
  Rb filled hollow core fibers},} in \enquote{CLEO/Europe and IQEC 2007
  Conference Digest,}  (Optical Society of America, 2007), p. IA\_4.

\bibitem{Ghosh2006}
S.~Ghosh, A.~Bhagwat, C.~Renshaw, S.~Goh, A.~Gaeta, and B.~Kirby,
  \enquote{Low-light-level optical interactions with rubidium vapor in a
  photonic band-gap fiber,} Physical Review Letters \textbf{97}, 023603 (2006).

\bibitem{Light2007}
P.~S. Light, F.~Benabid, F.~Couny, M.~Maric, and A.~N. Luiten,
  \enquote{{Electromagnetically induced transparency in Rb-filled coated
  hollow-core photonic crystal fiber.}} Optics Letters \textbf{32}, 1323--5
  (2007).

\bibitem{Uli}
U.~Vogl, C.~Peuntinger, N.~Y. Joly, P.~S. Russell, C.~Marquardt, and G.~Leuchs,
  \enquote{Atomic mercury vapor inside a hollow-core photonic crystal fiber,}
  Optics Express \textbf{22}, 29375--29381 (2014).

\bibitem{CLEO_2011}
W.~Zhong, B.~Heim, D.~Elser, C.~Marquardt, and G.~Leuchs,
  \enquote{{Polarization noise induced by photon-phonon interaction in
  hollow-core photonic crystal fibres },} in \enquote{Lasers and Electro-Optics
  Europe (CLEO EUROPE/EQEC), 2011 Conference on and 12th European Quantum
  Electronics Conference,}  (2011).

\bibitem{CLEO_US_2015}
W.~H. Renninger, H.~Shin, R.~O. Behunin, P.~Kharel, E.~Kittlaus, and P.~T.
  Rakich, \enquote{Stimulated forward brillouin scattering in hollow-core
  photonic crystal fiber,} in \enquote{CLEO: 2015, OSA Technical Digest,}
  (2015), SW4L.3.

\bibitem{Pang2010}
M.~Pang, H.~F. Xuan, J.~Ju, and W.~Jin, \enquote{{Influence of strain and
  pressure to the effective refractive index of the fundamental mode of
  hollow-core photonic bandgap fibers},} Optics Express \textbf{18}, 14041
  (2010).

\bibitem{Poustie1992}
A.~J. Poustie, \enquote{{Guided acoustic-wave Brillouin scattering with optical
  pulses},} Optics Letters \textbf{17}, 574 (1992).

\end{thebibliography}
\section{Introduction}
Light transmitted through optical fibers usually experiences interaction of the photons in the fiber core with thermally excited acoustic phonons in the fiber structure. The noise resulting from such photon-phonon interactions has been intensively investigated in standard fibers \cite{Shelby1985,Shelby1985a}, as well as in polarization maintaining fibers \cite{Nishizawa1995, Nishizawa1996} and solid-core photonic crystal fibers (PCFs) \cite{Laude2005,Shibata2006,Elser2006, Beugnot2007,McElhenny2011,Stiller_2011}. In all aforementioned cases, the acoustic vibrations in the fiber modulate the strain distribution in the silica fiber core, thus modulating the core refractive index and inducing polarization and phase noise to the transmitted light. This is known as guided acoustic-wave Brillouin scattering (GAWBS). Although the scattering is weak, the excess noise due to GAWBS can exceed the shot-noise level \cite{Elser2006}. This can be a problem if for example a fiber is used to generate squeezed states of light \cite{Shelby1986, Dong2008, Lorenz2001, Josip2007} or is used as a channel in continuous-variable quantum key distribution (QKD) experiments \cite{Lodewyck2007, Wittmann2010, Elser2007}. For this reason, the excess noise produced by such forward scattering in a fiber needs to be characterized thoroughly in order to calculate the noise floor.

Hollow-core photonic crystal fibers \cite{Knight1998, Cregan1999, RussellPCF} guide light in a central air hole surrounded by a periodic air-silica microstructure (see Fig. 1). The microstructure traps light in the core in specific wavelength ranges, called photonic bandgaps. This guiding mechanism offers a rich variety of novel applications. Since the light field has only marginal overlap with the fiber material, a high damage threshold in power delivery can be reached~\cite{Ouzounov2003}. Furthermore, the hollow core can be filled with gases or liquids enabling efficient light-matter interactions over much longer distances than in conventional gas cells. This allows for applications such as low-threshold gas lasers~\cite{Couny2007b}, optical sensors~\cite{Bykov2015}, particle guidance~\cite{Benabid2002}, photochemistry~\cite{Ana2013}, and ultra-violet light generation~\cite{KaFai2013}. Moreover, quantum-optical experiments can benefit from the enhanced light-matter interaction~\cite{Zhong2007}. In this direction, a hollow-core PCF filled with vapor in the core~\cite{Ghosh2006,Light2007}, offers the potential of generating squeezed light in resonant interactions~\cite{Zhong2007,Uli}.

The small overlap between light field and fiber structure might suggest hollow-core PCFs being essentially free from GAWBS noise, which would be an additional benefit for the aforementioned experiments. However, our first measurements show significant excess noise in hollow-core PCFs due to forward Brillouin scattering~\cite{CLEO_2011}. In addition, the stimulated regime of forward Brillouin scattering has attracted the researchers' interest~\cite{CLEO_US_2015}.

In this work, we show that forward scattering due to acoustic modes can still be observed in hollow-core PCFs. Our results reveal strong modulation due to acoustic vibrations of the photonic structure changing the effective refractive index of the optical mode guided in the hollow core.

\section{Experiment and results}
\begin{figure}[b!]
\centering\includegraphics[width=6cm]{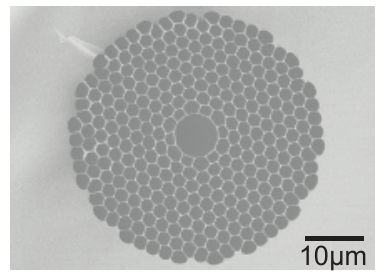}
\caption{Scanning electron micrograph (SEM) image of the core and photonic crystal structure of the hollow-core photonic crystal fiber used in our experiments (HC-800-02 Blazephotonics).}\label{fig:SemHCPCF}
\end{figure}

Our investigations are focused on the study of polarization noise in a commercially available hollow-core PCF (HC-800-02) from Blazephotonics with 8\,m length. As shown by the scanning electron microscopy (SEM) image in Fig.~\ref{fig:SemHCPCF}, it is a PCF with a hollow core made by removing 7 capillaries from the center during the stacking of the preform. The fiber has a core diameter of \unit{6.8}{\micro\meter}, a holey region diameter of \unit{41}{\micro\meter}, a silica cladding diameter of \unit{130}{\micro\meter} and a single layer acrylate coating with \unit{220}{\micro\meter} diameter. The air filling fraction in the holey region is more than 90\%.

\begin{figure}[t!]
\centering\includegraphics[width=9cm]{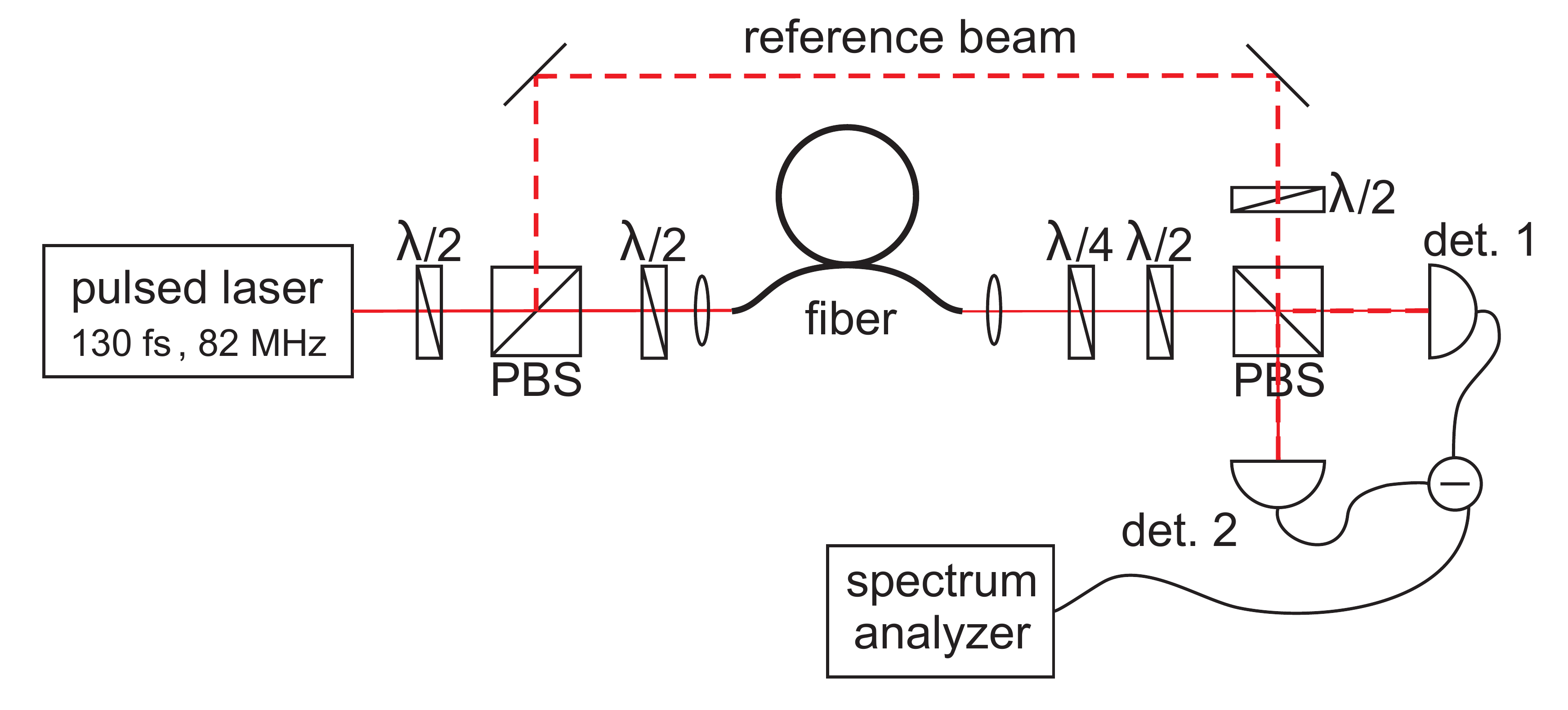}
\caption{Diagram of the experimental setup for detecting polarization noise in fibers. $\lambda/2$: half-wave plate; $\lambda/4$: quarter-wave plate; PBS: polarizing beam splitter; det.: detector (quantum noise limited).}\label{fig:Expsetup}
\end{figure}

In this paper, we focus on polarization noise induced by photon-phonon interactions. To prevent interferometric effects from reflections, a pulsed laser source is used. No intensity noise above shot-noise level is observed with a pulsed laser source (in contrast, we observed large intensity noise when continuous wave (cw) laser light is transmitted through a hollow-core PCF). The longitudinal mode spacing (pulse repetition frequency) of the mode-locked Ti:Sapphire laser (Spectra-Physics Tsunami) is \unit{82}{\mega\hertz} and its center wavelength of \unit{810}{\nano\meter} is located inside the bandgap of the hollow-core PCF. The setup shown in Fig.~\ref{fig:Expsetup} is used to measure the polarization noise spectrum in our hollow-core PCF. A half-wave plate and a polarizing beam splitter are used to divide the beam into two paths. One of the beams is launched into the fiber core via an aspheric lens and collimated at the fiber output via another aspheric lens with the same specification. A half-wave plate and a polarizing beam splitter after the fiber divide the beam into two parts of equal brightness. The two beams are detected by a pair of PIN-photodiode detectors whose amplified photo currents are then subtracted from each other. Polarization fluctuations of the light travelling through the fiber are converted into  oscillations of the photocurrent difference. We observe the generated sideband frequencies via a spectrum analyzer. Our hollow-core PCF is birefringent, thus the output beam is slightly elliptically polarized. To ensure a maximum extinction ratio for the detection, a half-wave plate at the fiber input side is used to properly align the input polarization and a quarter-waveplate at the output is inserted to compensate for the birefringence in the fiber and convert the output beam to linear polarization. The other output of the PBS (dashed line in Fig.~\ref{fig:Expsetup}) is used as a reference beam, indicating the shot-noise level by measuring the polarization noise of the shot-noise-limited laser source itself. 

\begin{figure}[htb]
\centering\includegraphics[width=9cm]{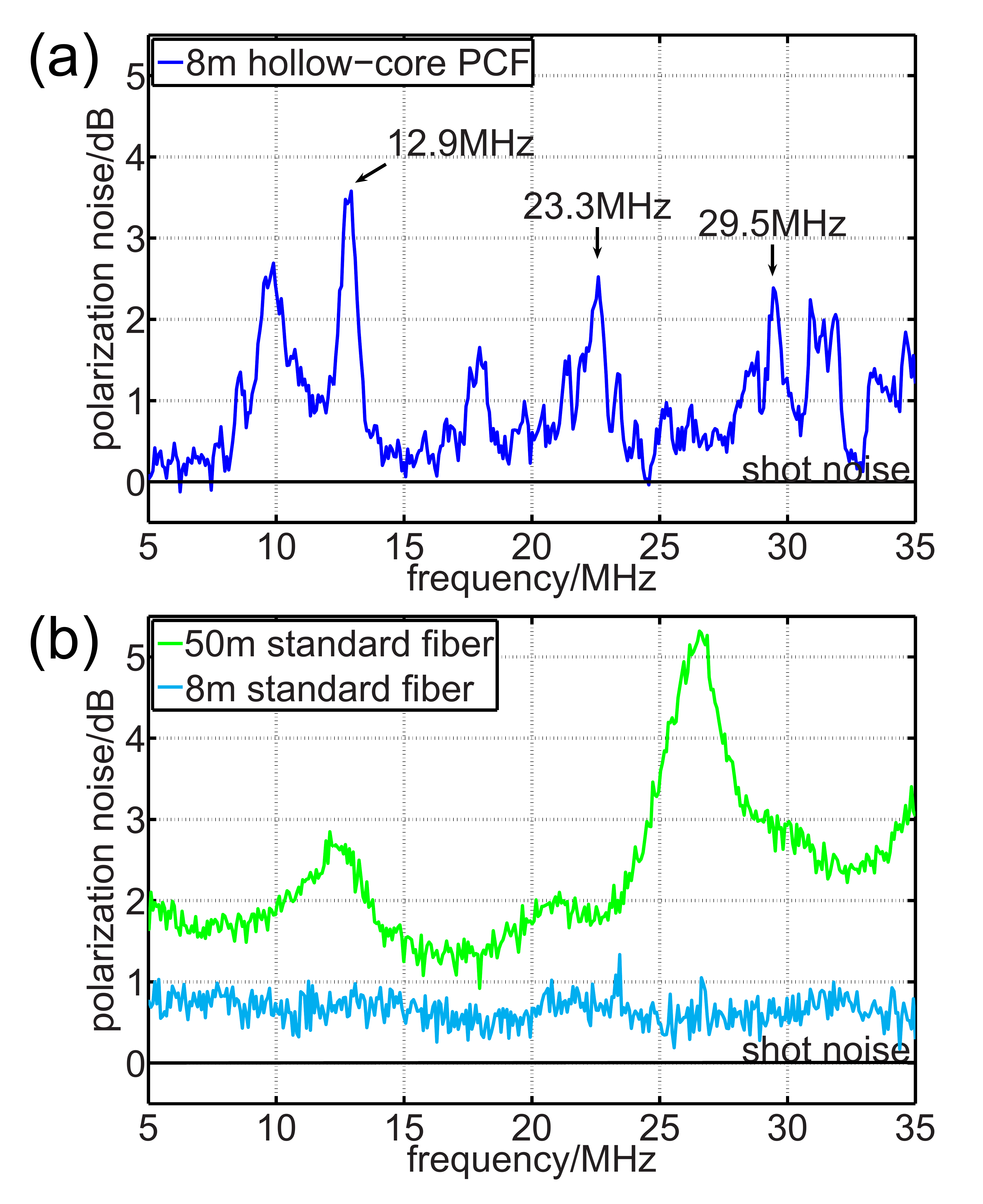}
\caption{Polarization noise spectrum relative to shot-noise for \unit{130}{\femto\second} pulses centered at \unit{810}{\nano\meter} with a \unit{82}{\mega\hertz} repetition rate and \unit{10}{\milli\watt} average power after fiber: (a) in \unit{8}{\meter} hollow-core photonic crystal fiber (Blazephotonics HC-800-02); (b) in \unit{8}{\meter} and \unit{50}{\meter} polarization maintaining fiber (Thorlabs FS-PM-4611). Three exemplary noise peaks in (a) are specified and marked with the value of the sideband frequency.}\label{fig:PolNoise}
\end{figure}

Fig.~\ref{fig:PolNoise}(a) shows the polarization noise spectrum of a pulsed laser beam after propagation through \unit{8}{\meter} of the hollow-core PCF. The traces are recorded at a resolution bandwidth of \unit{300}{\kilo\hertz}, a video bandwidth of \unit{300}{\hertz} and a 5 times averaging leading to a measurement accuracy of \unit{0.3}{\deci\bel}. The polarization noise spectra have been corrected for electronic noise of the photodetectors and are normalized to shot-noise (\unit{0}{\deci\bel} level in the figure). The laser is quantum noise limited above \unit{5}{\mega\hertz} and the detectors are linear up to \unit{35}{\mega\hertz}, thus our measurements are shown in the frequency range between \unit{5}{\mega\hertz} and \unit{35}{\mega\hertz}. To give a direct comparison of the polarization noise in hollow-core PCFs with the one in standard fibers, we measure the polarization noise spectrum in a standard fiber (Thorlabs FS-PM-4611). The optical power on the detectors (1\,mW) and the settings of the spectrum analyzer are the same as for the hollow-core PCF measurement. The acrylate coating has not been removed from all the fibers used in our measurement. The polarization noise spectra are shown in Fig.~\ref{fig:PolNoise}(b). For comparison, in an \unit{8}{\meter} standard fiber, the polarization noise structure is nearly vanishing in the quantum noise.
 
In 50\,m of standard fiber instead, polarization noise can be seen. In comparison, the measurement of the polarization noise in the hollow-core PCF reveals that in only 8\,m length of hollow core PCF, the intensity of the polarization noise peaks in the sidebands is comparable to the ones in 50\,m of standard fiber and higher than in 8\,m of standard fiber. This observation is surprising because light propagates mainly in the air core in a hollow-core PCF. Furthermore, as shown in Fig.~\ref{fig:PolNoise}, there are more polarization noise peaks in hollow-core PCF than in standard fiber in the measured frequency range. 

\section{Simulation and discussion}
As described by Shelby et al. \cite{Shelby1985,Shelby1985a}, the mixed torsional-radial (TR$_{2,m}$) modes of the transverse acoustic vibration of the cylindrical fiber are responsible for the polarization noise in standard fibers. These acoustic modes deform the fiber such as compressing the fiber along one axis and expanding the fiber along the orthogonal axis correspondingly. Such oscillations cause strain-induced density variations and hence refractive index variations in the fiber core which lead to birefringence variations. If one of the TR$_{2,m}$ modes is oscillating at \unit{45}{\degree} with respect to the input light polarization axis, the birefringence variations will induce polarization modulations at the frequency of the acoustic vibration. In other words, the sideband frequency of the observed polarization noise is the oscillation frequency of the TR$_{2,m}$ mode. 

\begin{figure}[t]
\centering\includegraphics[width=9cm]{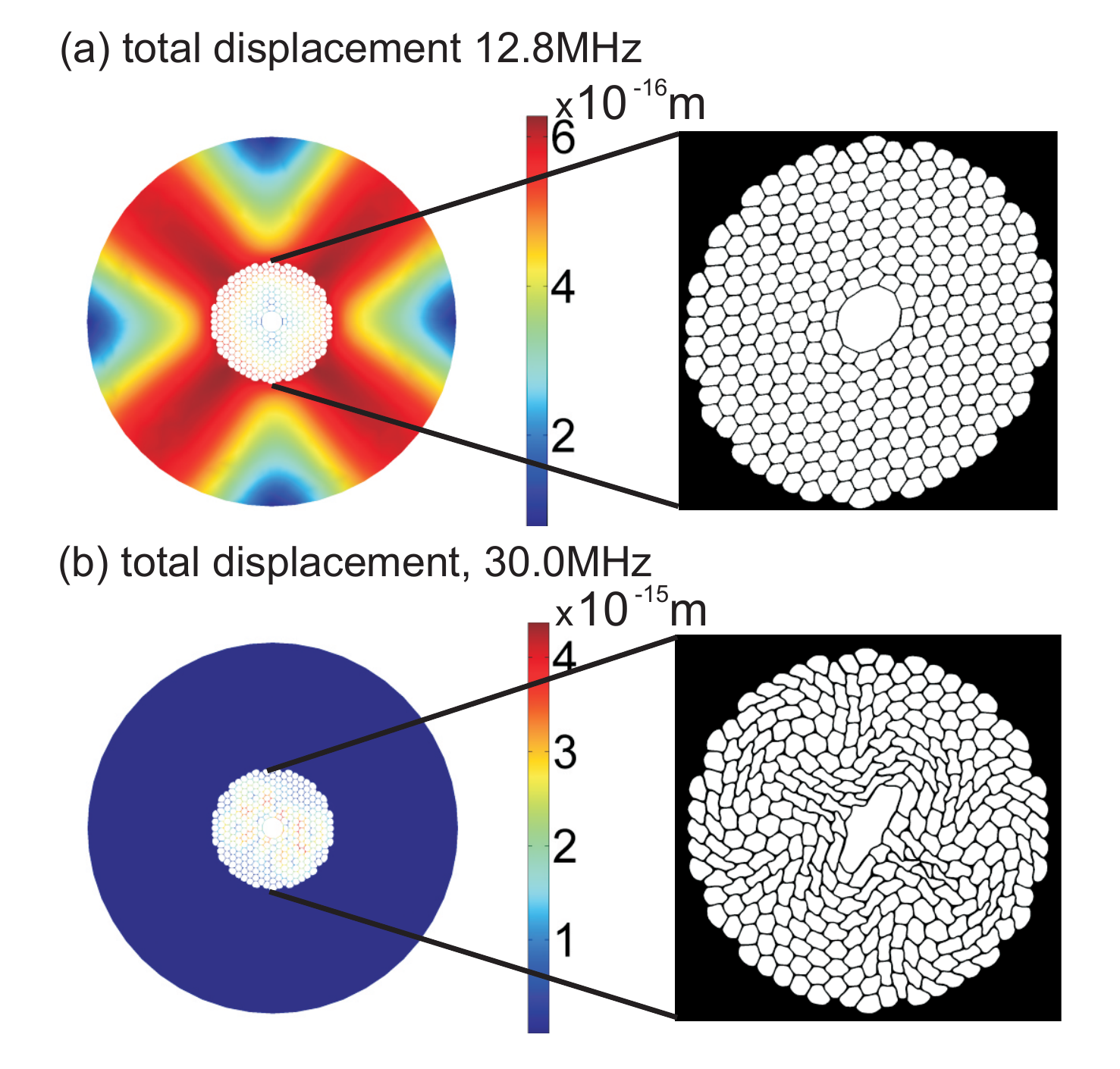}
\caption{Simulated acoustic eigenmodes in our hollow-core PCF. left: total displacement of the acoustic modes in the whole fiber cross section (red\,=\,large displacement, blue\,=\,small displacement); right: deformation of the photonic crystal structure due to acoustic vibrationv (deformation exaggerated by $2\times10^{9}$).} \label{fig:Simu}
\end{figure}

In our hollow-core PCF, since a large part of the laser light is propagating in the air core instead of in silica, the core refractive index modulation due to strain-induced density variations from acoustic vibration is strongly limited. However, the core refractive index modulation could be induced because of the photonic bandgap guidance in hollow-core PCF. It was presented by Pang et al. \cite{Pang2010} that deformations of the photonic bandgap structure change the effective refractive index sensed by the light propagating in the air core in a steady state situation. In our assumption, the acoustic vibrations in the hollow-core PCF cause both dynamic geometry deformations and material index modifications of the bandgap structure which effectively leads to birefringence modulations.

\begin{figure}[ht]
\centering\includegraphics[width=11cm]{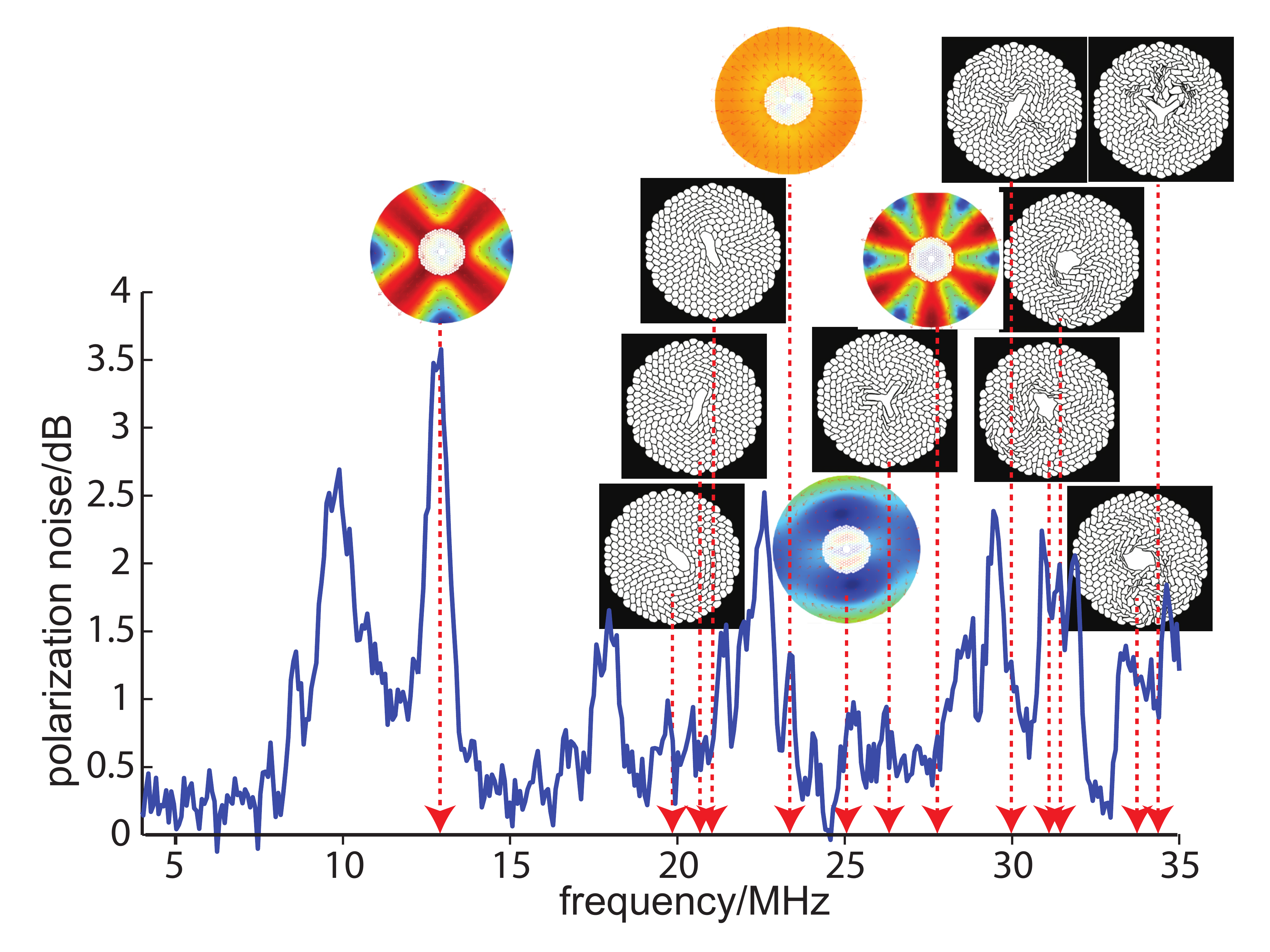}
\caption{Correspondence of the experimental polarization noise peaks and the simulated acoustic modes of the hollow-core PCF. The color coded pictures show the silica cladding vibration modes (red=large displacement, blue=small displacement). The black and white pictures show the deformation of fiber via photonic crystal structure vibration modes.} \label{fig:Match}
\end{figure}
 
In order to verify this assumption, we calculated the transverse acoustic modes which are resonant (eigenmodes) of the hollow-core PCF and compared their resonant frequency with our observation. As the structure of a realistic hollow-core PCF is too complex to be calculated analytically, we used the finite element method (COMSOL Multiphysics) to solve for the acoustic eigenmodes of the fiber structure. We took a high resolution SEM image of the photonic bandgap structure of our fiber and manually transformed the image into the model geometry.

We performed an eigenfrequency analysis in a two-dimensional plane strain model. The fiber deformation due to acoustic vibrations was normalized as described by Shelby et al. \cite{Shelby1985} such that the total vibration energy for each acoustic mode in the fiber must be equal to $k_{B}T$. The simulations result in the acoustic eigenfrequencies as well as the displacement for each vertex of the finite elements mesh. Fig.~\ref{fig:Simu} exemplarily shows two of the acoustic modes we obtained from the simulation. We observe two different types of acoustic modes. The first type (Fig.~\ref{fig:Simu}(a)) alters the whole structure and the second type of acoustic modes (Fig.~\ref{fig:Simu}(b)) is limited to the photonic structure. We distinguish them as they lead to a different physical behavior with respect to the optical mode. 

The frequency of the first simulated acoustic mode vibrating at \unit{12.8}{\mega\hertz} (Fig.~\ref{fig:Simu}(a)) fits well with the experimentally observed polarization noise peak at a side band frequency of \unit{12.9}{\mega\hertz}. This mode is similar to a TR$_{2,0}$ mode in a standard fiber. A rough calculation of the elasto-optic coefficient~\cite{Stiller_2011} shows that this noise peak, as well as the one at 23.3\,MHz, is caused by the optical mode overlapping with the acoustic mode in the inner ring of the photonic structure. In other words, it is the same physical effect as in solid-core PCF as light leaks into the silica material of the microstructure.
\\The influence of the photonic crystal structure on TR$_{2,m}$ modes in solid-core PCFs has been investigated by McElhenny et al. \cite{McElhenny2011}. They state that usually for lower order acoustic modes, the overall acoustic mode shape is barely altered by the air holes. 
The photonic crystal structure is deformed by the vibration of the silica cladding. In Fig.~\ref{fig:Simu}(a) right, a snapshot of the exaggerated deformation of the photonic crystal structure deformed by the vibration is shown.
\\
The second type of acoustic vibration modes, photonic crystal structure vibration modes, is shown in Fig.~\ref{fig:Simu}(b). The calculated eigenfrequency of \unit{30.0}{\mega\hertz} agrees well with the experimentally measured polarization noise peak at a sideband frequency of \unit{29.5}{\mega\hertz}. Here, only the photonic crystal structure vibrates and the outer silica cladding remains motionless. By these kind of acoustic vibrations, additional polarization noise peaks are induced and cause the polarization noise spectrum of light traveling through a hollow-core PCF to be more complex than the one in a standard fiber.  

In Fig.~\ref{fig:Match}, we compare the experimentally measured noise spectrum with our simulated acoustic vibration modes. Most of the simulated acoustic modes with vibration frequencies below \unit{35}{\mega\hertz} are shown in the figure with a fiber deformation picture and an arrow marking the eigenfrequency. The corresponding sideband frequencies of the polarization noise and the simulated acoustic frequencies of the eigenmodes lie within an accuracy of 0.5\,MHz. One encounters both modes stemming from vibrations of the whole fiber (color coded pictures) and types of modes stemming from the photonic crystal structure vibrations alone (black and white pictures). $13$ out of $16$ of the simulated acoustic modes with vibration frequencies below \unit{35}{\mega\hertz} are shown in the figure; only for three acoustic modes corresponding peaks could not be found in
the experimental noise spectrum. For a continuous wave laser, the measured spectrum would be the result of the beating between the forward scattered (depolarized) light and the unscattered (input) light. The center frequencies of the peaks in the polarization noise spectrum thus correspond to the frequencies of the acoustic modes that depolarized the input light. In our experiment, a mode-locked laser has been used in order to prevent cavity effects in the fiber. Considering this fact,
the scattered light from one cavity mode of the mode-locked laser not only beats with its original mode, but also beats with the other cavity modes which leads to a series of noise peaks in the spectrum that are repeated at the laser repetition rate \cite{Poustie1992}. For instance, the scattered light of one cavity mode, which is shifted by a \unit{92}{\mega\hertz} acoustic vibration, beats with the adjacent cavity mode which is shifted by a laser repetition rate of \unit{82}{\mega\hertz} compared to the first cavity mode, and results in the polarization noise at sideband frequency of $\unit{92}{\mega\hertz}-\unit{82}{\mega\hertz}=\unit{10}{\mega\hertz}$. We assume this to be the reason why a few of the measured noise peaks in Fig.~\ref{fig:Match} do not have corresponding simulated acoustic modes at the same frequencies.


\section{Summary}
To summarize, we have reported on the first measurement of guided acoustic forward scattering in a hollow-core PCF. In a quantum noise limited detection, we have observed polarization noise in hollow-core PCF using a pulsed laser. Several forward scattering modes have been observed and identified by comparing the experimental spectrum with the calculated eigenfrequencies of fiber vibration obtained using the finite element method. The observed modes are shown to be whole fiber torsional-radial (TR) modes, as well as vibrational modes of the photonic bandgap structure alone. A newly discovered forward scattering mechanism in optical fibers, which can be attributed to dynamic geometric variation of the photonic bandgap structure instead of density modulation of the fiber core, is presented. The excess polarization noise in our hollow core PCF has been characterized in shot-noise limited measurements. Our results are important for further applications of this fiber in quantum optical experiments.

\end{document}